\documentclass[preprint,aps,12pt,preprintnumbers,eqsecnum,nofootinbib]{revtex4}
\usepackage{graphicx}
\usepackage{subfigure}

\usepackage{color}
\usepackage{amssymb,amsmath}

\newcommand{\beq}{\begin{equation}}
\newcommand{\enq}{\end{equation}}

\begin{document}
%
%
\title{\vspace*{0.5in} 
Acausality from a Dark Sector
\vskip 0.1in}
\author{Christopher D. Carone}\email[]{cdcaro@wm.edu}

\affiliation{High Energy Theory Group, Department of Physics,
College of William and Mary, Williamsburg, VA 23187-8795}
\date{December 2013}

\begin{abstract}
Solutions to the hierarchy problem that require partners for each standard model particle often require that these states 
live at or above the electroweak scale, to satisfy phenomenological bounds.  Partners to possible dark sector particles may
be significantly lighter, due to the assumed weakness of the couplings between the dark and visible sectors. Here we 
consider the possibility that a dark sector might include light Lee-Wick particles.   We present the formulation of a 
theory in which a dark photon and its Lee-Wick partner have kinetic mixing with hypercharge.  We point out that the 
Lee-Wick partner of the dark photon will lead to an apparent violation of causality at small distance scales that might be 
discerned in low-energy experiments.
\end{abstract}
\maketitle

\section{Introduction}\label{sec:intro}
The strong indirect evidence for the existence of dark matter has motived great interest in the possibility of an 
associated sector of new particles that communicates only weakly with the standard model.   Such a sector may
include a spontaneously broken U(1) gauge symmetry under which a dark matter candidate is charged.  If the
scale of spontaneous symmetry breaking in the dark sector is below a GeV, then one can potentially account for the 
cosmic ray electron and positron excess observed in the PAMELA~\cite{pamela}, Fermi-LAT~\cite{fermilat} and other 
experiments~\cite{ams02}:  dark matter may annihilate to the sub-GeV scale dark photons, which subsequently decay to 
leptons~\cite{nima}.   Decays to protons and anti-protons (for which there is no observed excess) are not kinematically 
allowed.  Assuming this mechanism, the possibilities for additional particles and symmetries of the dark sector are not 
substantially constrained.

One possibility for the additional particle content of the dark sector are partner particles associated with a 
mechanism that addresses the hierarchy problem.  In the visible sector, such partner particles are 
typically above the electroweak or TeV scale, given stringent collider and indirect constraints.  However, the assumed weakness 
of the couplings between the dark and visible sectors allows the dark-sector partners to be substantially lighter.    For example, in
supersymmetric models, each standard model particle has a partner with opposite spin statistics that 
cancels the quadratic divergence in the Higgs boson squared mass.  The masses of these partners are input via terms that 
softly break the supersymmetry; the value of the soft-breaking parameters are free, unless one specifies a model for the 
supersymmetry breaking sector.   If the minimal supersymmetric standard model is extended by a dark sector (for example, 
see Ref.~\cite{dsusy}) nothing forbids the mass scale of the dark superpartners from being substantially smaller 
than that of the visible sector.   This may lead to new experimental avenues for the discovery of supersymmetry~\cite{dsusy}.  
On the other hand, in models where the hierarchy problem is solved via a warped extra dimension, it is possible that a dark 
sector lives in its own warped ``throat"~\cite{warpdark} with a distinct infrared scale.   In such models, towers of 
dark-sector Kaluza-Klein (KK) excitations may begin below the electroweak scale, but have evaded detection due to the 
weakness of their couplings.

Here we suggest a different possibility, namely, that the dark sector may include light Lee-Wick (LW) partners.   The 
Lee-Wick Standard Model (LWSM) has been proposed as a solution to the hierarchy problem~\cite{lwsm} and has received
attention in the recent literature~\cite{lwrecent}; dark LW particles have been discussed recently in the context of a model of flavor in
Ref.~\cite{gab}. In the LWSM, each standard model field is given a higher-derivative 
kinetic term which introduces an additional pole in the propagator, corresponding to a partner particle.  Eliminating the 
higher-derivative term using an auxiliary field approach (which we will illustrate in the next section), one finds that the LW 
partner has wrong-sign kinetic and mass terms; this implies that the LW state must have negative norm.   Lee, Wick~\cite{lworig} and 
others~\cite{clop} showed that at least up to one loop the presence of negative norm states does not lead to a violation of the 
unitarity of the $S$-matrix or Lorentz invariance provided that (i) the LW particles are unstable and (ii) loop integrals are evaluated 
with a carefully chosen pole prescription.  It has been speculated that a prescription that preserves unitarity and Lorentz invariance
exists beyond one loop.  Cancellations between diagrams involving ordinary particles and diagrams involving LW partners eliminates the 
unwanted quadratic  divergences. The result is a theory with only logarithmic sensitivity to high scales.  

No phenomenological constraint prevents the LW partners of a sub-GeV scale dark sector from being much lighter than 
the partners of the visible sector particles.   Since there are no known examples of theories that predict the LW mass 
spectrum {\em ab initio}, this is simply a parametric possibility.  However, it is an interesting one for the following 
reason: Lee-Wick theories have been argued to be the only known consistent examples of acausal 
quantum field theories~\cite{coleman,kuti,emergent}.  One indication of acausality is the presence of ``wrong displaced 
vertices:"  decay products from a LW resonance move towards a primary production vertex rather than away from it~\cite{coleman,emergent,schat}.   
This phenomenon has been discussed in the context of Large Hadron Collider (LHC) searches for Lee-Wick particles in Ref.~\cite{schat}.   
Here we point out that the potential lightness of dark-sector LW particles may make such a phenomenon accessible at 
low-energy fixed target or collider experiments as well.  In analogy to the examples of dark sector supersymmetry or extra dimensions, 
the possibility of light LW partners in a dark sector may provide new avenues for the experimental discovery of LW particles 
and the observation of their unusual acausal phenomenology.

Given the motivation for dark photons that we described earlier, we focus here on the possibility that the LW partner 
to the dark photon is light.  We assume for simplicity that all other LW partners are heavy and decoupled from the 
effective theory.    In this letter we discuss the formulation of the theory and some of its phenomenological consequences. 

\section{Framework} \label{sec:model}

We consider a theory with higher-derivative kinetic terms as well as kinetic mixing between a dark Abelian gauge field 
$\hat{A}_D^\mu$ and the hypercharge gauge field $\hat{B}^\mu$.  We use hats to denote fields in the higher-derivative 
formulation of the theory.  After electroweak symmetry breaking, $\hat{B}^\mu$ mixes with the third component 
of the weak SU(2) gauge multiplet, $\hat{W}_3^\mu$; hence, we begin by considering the pure gauge field part of 
the Lagrangian that is quadratic in $\hat{A}^\mu \equiv (\hat{A}_D^\mu, \hat{B}^\mu, \hat{W}_3^\mu)^T$.   In terms of 
this column vector of fields, the quadratic portion of the Lagrangian may be written
\begin{equation}
{\cal L}_2 = -\frac{1}{4} \hat{F}_{\mu\nu}^T K_2 \hat{F}^{\mu\nu} + \frac{1}{2 \Lambda^2} \partial^\mu\hat{F}_{\mu\nu}
K_4 \partial^\lambda {\hat{F}_\lambda}^{\,\,\,\nu} + \frac{1}{2} \hat{A}^T_\mu M^2 \hat{A}^\mu \,\, ,
\label{eq:start}
\end{equation}
where $\Lambda$ is the Lee-Wick scale and $\hat{F}_{\mu\nu}=\partial_\mu \hat{A}_\nu-\partial_\nu \hat{A}_\mu$.
The kinetic terms that are quadratic order in derivatives include the conventional kinetic mixing term via
\begin{equation}
K_2 = \left(\begin{array}{ccc} 1 & c & 0 \\ c & 1 & 0 \\ 0 & 0 & 1 \end{array}\right) \,\, ,
\label{eq:k2}
\end{equation}
where $c$ is the kinetic mixing parameter, which we will assume is much smaller than one.  The higher-derivative 
term in Eq.~(\ref{eq:start}) can also introduce further mixing between $\hat{A}_D$ and $\hat{B}$.
We adopt the simplest form that is consistent with the assumption that the Lee-Wick partners of the 
visible-sector particles are decoupled, namely
\begin{equation}
K_4 = \left(\begin{array}{ccc} 1 & 0 & 0 \\ 0 & 0 & 0 \\ 0 & 0 & 0 \end{array}\right) \,\, .
\end{equation}
In this minimal formulation, all communication between the dark and visible sectors is controlled by the parameter $c$.
Finally, the mass squared matrix $M^2$ in Eq.~(\ref{eq:start}) includes all masses that arise via spontaneous 
symmetry breaking in both the dark sector and the visible sector.  Since we assume no Higgs fields that are charged
under both the visible and dark gauge groups, this matrix takes the form
\begin{equation}
M^2 =\frac{v^2}{4} \, \left(\begin{array}{ccc} \xi & 0 & 0 \\ 0 & g_Y^2 & -g \, g_Y \\ 0 & -g \, g_Y & g^2 \end{array}\right) \,\, ,
\end{equation}
where $g$ and $g_Y$ are the weak SU(2) and hypercharge gauge couplings, respectively.   The parameter $\xi$ is given by
$g_D^2 v_D^2/v^2$, where $g_D$ is the dark gauge coupling and $v_D$ is the scale at which the U(1)$_D$ is 
spontaneously broken.  We assume that  $v_D$ is at or below the GeV scale, so that $\xi \ll 1$.

We wish to ascertain the spectrum and couplings of the dark sector states that are implied by Eq.~(\ref{eq:start}).  We first
perform the change of basis $ \hat{A} = R_1 \hat{A}_1 $, where
\begin{equation}
R_1 = \left(\begin{array}{ccccc} 1/\sqrt{1-c^2} & & 0 & &  0 \\ -c/\sqrt{1-c^2} & & c_w & & -s_w \\ 0 
& & s_w & & c_w \end{array}\right) \,\, , 
\label{eq:r1def}
\end{equation} 
with $s_w = \sin\theta_w$ and $c_w=\cos\theta_w = g/\sqrt{g^2+g_Y^2}$.  The effect of this redefinition on the 
matrices in Eq.~(\ref{eq:start}) is
\begin{equation}
R_1^T K_2 R_1 = \openone  \,\, ,
\label{eq:r1k2}
\end{equation}
\begin{equation}
R_1^T K_4 R_1 = \frac{1}{1-c^2} K_4 \,\, ,
\label{eq:r1k4}
\end{equation}
and 
\begin{equation}
R_1^T M^2 R_1 = \left(\begin{array}{ccccc}
({m_D}^2 +c^2 s^2_w {m_Z}^2)/(1-c^2) & & 0 & & c \,s_w {m_Z}^2 /\sqrt{1-c^2} \\
0 & & 0 & & 0 \\
c\, s_w {m_Z}^2/ \sqrt{1-c^2}   & & 0 & & {m_Z}^2 \end{array} \right) \,\,\, .
\label{eq:r1mass}
\end{equation}
In Eq.~(\ref{eq:r1mass}), $m_D^2 = g_D^2 v_D^2/4$ and $m_Z^2 =(g^2+g_Y^2) v^2/4$  are the mass 
eigenvalues of the dark gauge boson and the $Z$ boson, respectively, in the $c \rightarrow 0$, 
$\Lambda \rightarrow \infty$ limit.  By studying the form of the covariant derivative, it is straightforward to 
confirm that the second component of the column vector $\hat{A}_1$, which corresponds to the vanishing 
row and column of Eq.~(\ref{eq:r1mass}), couples to matter proportional to $e\, (Y+T^3)$, where 
$e= c_w \, g_Y = s_w \, g$ and $Y$ ($T^3$) is the generator of hypercharge (the third component of weak isospin).  
Hence, we identify the second component of $\hat{A}_1$ as the photon field.  Since Eq.~(\ref{eq:r1k4}) provides 
no higher-derivative kinetic term for this field, we may drop the hat;  we recover an ordinary photon with the usual 
couplings and no ({\em i.e.}, a completely decoupled) Lee-Wick partner.  Hence, we ignore the QED sector of the 
theory and focus on the remaining fields, the first and third component of $\hat{A}_1$, which we will call
$\hat{A}_{1D}$ and $\hat{Z}_1$, respectively.

We next rid ourselves of the $\hat{A}_{1D}$ higher-derivative kinetic term by employing the standard auxiliary field 
approach~\cite{lwsm}
\begin{equation}
{\cal L}_{aux} = - \frac{1}{4} \hat{F}_{1D\mu\nu} \hat{F}^{\mu\nu}_{1D} 
- \frac{1}{4} \hat{Z}_{1\mu\nu} \hat{Z}_1^{\mu\nu}
-\frac{1}{2} \Lambda^2 (1-c^2) \tilde{A}^\mu_{1D} \tilde{A}_{1D \mu}
+\partial ^\mu \tilde{A}_{1D}^\nu \hat{F}_{1D\mu\nu} + {\cal L}_{SSB} \,\, ,
\label{eq:auxlag}
\end{equation}
where ${\cal L}_{SSB}$ represents the mass terms originating from spontaneous symmetry breaking ({\em i.e.} from 
the non-vanishing two-by-two sub-block of Eq.~(\ref{eq:r1mass}).  Substitution of the equation of motion for the 
auxiliary field $\tilde{A}^\mu_{1D}$ into Eq.~(\ref{eq:auxlag}) reproduces the higher-derivative Lagrangian defined by
Eqs.~(\ref{eq:r1k2}), (\ref{eq:r1k4}), and (\ref{eq:r1mass}).  Now defining a shifted field $A_{1D}$ by
\begin{equation}
\hat{A}_{1D} = A_{1D} + \tilde{A}_{1D} \,\,\, ,
\label{eq:shift}
\end{equation}
and $\hat{Z}_1= Z_1$ (with no shift), Eq.~(\ref{eq:auxlag}) may be rewritten as
\begin{equation}
{\cal L}_{LW} = 
- \frac{1}{4} F_{1D\mu\nu} F^{\mu\nu}_{1D} 
- \frac{1}{4} Z_{1\mu\nu} Z_1^{\mu\nu}+ \frac{1}{4} \tilde{F}_{1D\mu\nu} \tilde{F}^{\mu\nu}_{1D} 
-\frac{1}{2} \Lambda^2 (1-c^2) \tilde{A}^{\mu}_{1D} \tilde{A}_{1D \mu} + {\cal L}_{SSB}
\label{eq:lwform}
\end{equation}
where
\begin{equation}
{\cal L}_{SSB} = \frac{1}{2} A_s^T M^2_{SSB} A_{s}
\label{eq:mssbterm}
\end{equation}
where, after the shift, we define $A_s=(\tilde{A}_{1D},\, A_{1D},\, Z_1)$ and
\begin{equation}
M^2_{SSB} =\left(\begin{array}{ccc}
 ({m_D}^2 +c^2 s^2_w {m_Z}^2)/(1-c^2) & ({m_D}^2 +c^2 s^2_w {m_Z}^2)/(1-c^2) &  c \,s_w {m_Z}^2 /\sqrt{1-c^2} \\
({m_D}^2 +c^2 s^2_w {m_Z}^2)/(1-c^2) & ({m_D}^2 +c^2 s^2_w {m_Z}^2)/(1-c^2) &  c \,s_w {m_Z}^2 /\sqrt{1-c^2} \\
c \,s_w {m_Z}^2 /\sqrt{1-c^2} & c \,s_w {m_Z}^2 /\sqrt{1-c^2} & {m_Z}^2 \end{array}\right) \,\, . \,\,
\label{eq:msqexp}
\end{equation} 
The mass matrix for the dark gauge boson, its Lee-Wick partner and the $Z$ boson is obtained by
adding the fourth term of Eq.~(\ref{eq:lwform}) to Eq.~(\ref{eq:mssbterm}).  The resulting matrix can be 
diagonalized by a sequence of three rotations:  (i) a unitary rotation acting on the $2$-$3$ block, (ii)
a symplectic rotation acting on the $1$-$3$ block, and (iii) a symplectic rotation acting on the $1$-$2$ block.
By a symplectic rotation, we mean a rotation of the form
\begin{equation}
\left(\begin{array}{cc} \cosh\theta & \sinh\theta \\ \sinh\theta & \cosh\theta \end{array}\right)
\end{equation}
which preserves the normalization and relative sign difference between the kinetic terms of an ordinary 
and a Lee-Wick field.  Since $c \ll 1$, the $2$-$3$ and $1$-$3$ rotation angles are small, while the 
$1$-$2$ angle generically is not.   Evaluating the small angles to lowest order in a series expansion in the 
kinetic mixing parameter $c$, as well as the small mass ratios $m_D^2/m_Z^2$  and $\Lambda^2/m_Z^2$, we find
\begin{equation}
A_s = R_2 \, A_2 \,\, ,
\end{equation}
where the mass eigenstate fields are $A_2 \equiv (\tilde{A}_{D} , A_{D} , Z)$ and
\begin{equation}
R_2 \approx \left(\begin{array}{ccc} \cosh\theta & \sinh\theta & -c \, s_w \\
\sinh\theta & \cosh\theta & c \, s_w \\
-c \, s_w e^{\theta} & -c \, s_w e^{\theta} & 1 \end{array} \right)  \,\, ,
\end{equation}
where $\tanh2\theta \approx 2 m_D^2 / (\Lambda^2-2 m_D^2)$.

We now determine the couplings of the dark photon and its LW partner to ordinary standard model matter.
To do so, we begin with the covariant derivative in our original basis, retaining only the neutral gauge fields:
\begin{equation}
D^\mu  = \partial^\mu  - i g_D Y_D \hat{A}_D^\mu - i g_Y Y \hat{B}^\mu -i g\, T^3 \hat{W_3}^\mu  \,\,\, .
\end{equation}
After the field redefinition specified by $R_1$ in Eq.~(\ref{eq:r1def}) and the shift of Eq.~(\ref{eq:shift}),
\begin{equation}
D^\mu = \partial^\mu -i \frac{g_D Y_D - c g_Y Y}{\sqrt{1-c^2}} (A_{1D}^\mu + \tilde{A}_{1D}^\mu)
-i e\, (Y+T^3) A^\mu  -i \frac{e}{s_w c_w} (c_w^2 T^3-s_w^2 Y) Z_1^\mu \,\,\, .
\end{equation}
Ignoring the photon, this corresponds to the basis previously called $A_s$, with the non-diagonal mass matrix given by
Eqs.~(\ref{eq:lwform})-(\ref{eq:msqexp}).   Rotating to the mass eigenstate basis, again working
only to lowest order in the kinetic mixing parameter and small mass scale ratios,
\begin{equation}
D^\mu = \partial^\mu -i e\, Q A^\mu- i e^{\theta} (g_D Y_D - c \, e Q c_w) (\tilde{A}_D^\mu + A_D^\mu)
- i \frac{e}{s_w c_w} (T^3 -s_w^2 Q)  Z^\mu  \,\,\,.
\end{equation}
The field redefinitions that we have employed to obtain canonical and diagonal quadratic terms (aside from LW signs) 
have had the following cumulative effect: The LW dark photon has the same couplings to standard model particles as the 
dark photon itself, proportional to $\epsilon \, e\,  Q$, where we define $\epsilon \equiv c\, c_w \exp\theta$.
Corrections to the $Z$-boson couplings occur at higher order in the small parameters, and will be 
negligible for the parameter choices relevant to our subsequent discussion.

\section{Divergences}
In this section, we verify that the presence of kinetic mixing does not alter the cancellation of quadratic 
divergences in gauge boson loops.  For this purpose, we ignore masses generated
via spontaneous symmetry breaking (which do not affect the ultraviolet behavior of the theory) and 
focus on the equivalent unbroken U(1)$_D \times$ U(1)$_Y$ LW gauge theory.  Hence, we start with a 
simpler version of Eq.~(\ref{eq:start}), defining in this case $\hat{A}^\mu \equiv (\hat{A}_D^\mu, \hat{B}^\mu)^T$ 
and setting $M^2=0$. Since we wish now to retain LW partners for both the dark and hypercharge gauge 
bosons, we assume a completely general form for the matrix $K_4$.  We perform the first field redefinition 
$\hat{A} = R_1 \hat{A}_1$ with
\begin{figure}[t]
\includegraphics[width = 0.495\textwidth]{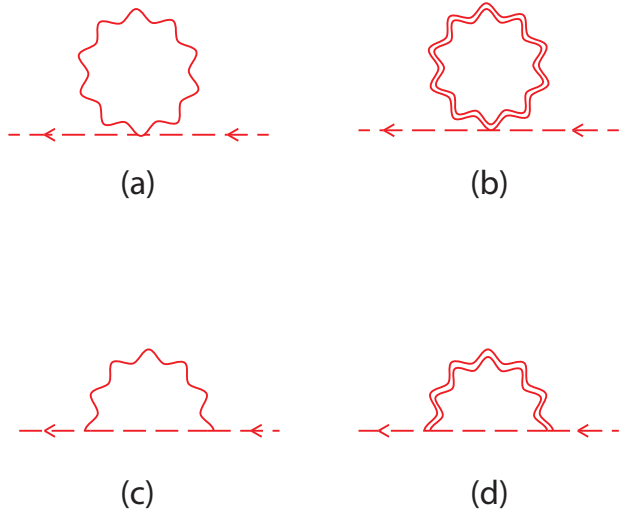}
\caption{\label{fig:lwloop} Contributions to the Higgs doublet self energy from the U(1)$_D \times$U(1)$_Y$
gauge fields (wavy lines) and LW partners (double wavy lines).}
\end{figure}    
\begin{equation}
R_1 = \left(\begin{array}{ccc} 1/\sqrt{1-c^2} & \,\,\, &0 \\ -c/\sqrt{1-c^2} & &1 \end{array} \right) \,\,\, ,
\end{equation}
which again diagonalizes the matrix $K_2$.   Unlike Eq.~(\ref{eq:r1k4}) however, $R_1^T K_4 R_1 \equiv K_4'$ is, in the 
present case, a completely general two-by-two, real, symmetric matrix.  Thus we write
\begin{equation}
K_4' = O^T K_D \, O \,\,\, ,
\end{equation}
where $O$ is an orthogonal matrix and $K_D$ is diagonal.  We can now state the generalizations of the auxiliary field
and LW Lagrangians, including only terms quadratic in the fields,
\begin{equation}
{\cal L}_{aux} = - \frac{1}{4} \hat{F}^{T}_{1\mu\nu} \hat{F}_1^{\mu\nu}
-\frac{1}{2} \Lambda^2  \tilde{A}^T_\mu K_D^{-1}\tilde{A}^{\mu}
+\partial^\mu \tilde{A}^{T \, \nu} O \hat{F}_{1\mu\nu} \,\,\,,
\end{equation}
\begin{equation}
{\cal L}_{LW} = - \frac{1}{4}  F^{\prime\, T}_{\mu\nu} F^{\prime \mu\nu}
+\frac{1}{4} \tilde{F}^{T}_{\mu\nu} \tilde{F}^{\mu\nu}
-\frac{1}{2} \Lambda^2  \tilde{A}^T_\mu K_D^{-1}\tilde{A}^{\mu} \,\,\, ,
\label{eq:lw2}
\end{equation}
which are related by the shift $\hat{A}_1 = A' + O^T \tilde{A}$.   Hence, fields in the original basis are related to
the fields in Eq.~(\ref{eq:lw2}) by
\begin{equation}
\hat{A} = R_1 ( A'+ O^T \tilde{A}) \,\,\, .
\label{eq:newshift}
\end{equation}

Couplings of the fields $A'$ and $\tilde{A}$ to the Higgs doublet field $H$ arise through the kinetic term
$D^\mu H^\dagger D_\mu H$, where $D_\mu H= [\partial_\mu -(i/2) g^T \hat{A}_\mu] H$.  Since $\hat{A}$ is
a column vector, we have defined $g^T = (0, g_Y)$.  The relevant couplings are given by
\begin{equation}
{\cal L} \supset (g^T R_1 A'_\mu + g^T R_1 O^T \tilde{A}_\mu) (\frac{i}{2}H^\dagger \partial^\mu H + \mbox{ h.c.}) 
+ \frac{1}{4} (g^T R_1 A'_\mu + g^T R_1 O^T \tilde{A}_\mu)^2 H^\dagger H \, .
\end{equation}
We may now evaluate the four self-energy diagrams shown in Fig.~\ref{fig:lwloop}, with massless gauge field
propagators written in Feynman gauge and defining $\tilde{M}^2 = \Lambda^2 K_D^{-1}$:
\begin{equation}
-i \Sigma(0)_a = \frac{1}{4} g^T R_1 R_1^T \, g \int \frac{d^d k}{(2 \pi)^d} \frac{d}{k^2}  \,\,\, .
\label{eq:normala}
\end{equation}
\begin{equation}
-i \Sigma(0)_b  = -\frac{1} {4} g^T R_1 \, O^T \left[ \int \frac{d^d k}{(2 \pi)^d} [ (d-1)[k^2-\tilde{M}^2]^{-1} - (\tilde{M}^2)^{-1} ] 
\right] O \, R_1^T g \,\,\, ,
\label{eq:normalb}
\end{equation}
\begin{equation}
-i \Sigma(0)_c = -\frac{1}{4} g^T R_1 R_1^T \, g \int \frac{d^d k}{(2 \pi)^d} \frac{1}{k^2} \,\,\, ,
\label{eq:normalc}
\end{equation}
\begin{equation}
-i \Sigma(0)_d=  -\frac{1} {4} g^T R_1 \, O^T \left[ \int \frac{d^d k}{(2 \pi)^d} (\tilde{M}^2)^{-1} \right] O \, R_1^T g \,\,\,.
\label{eq:normald}
\end{equation}
These quantities give the shift in the physical Higgs doublet mass in the absence of a tree-level scalar potential, the 
generalizations of Eqs.~(33a)-(33d) in Ref.~\cite{lwsm}.  Quartic divergences again cancel between 
Eqs.~(\ref{eq:normalb}) and (\ref{eq:normald});  quadratic divergences in the sum of Eqs.~(\ref{eq:normala}) 
and (\ref{eq:normalc}) cancel those in Eq.~(\ref{eq:normalb}).

It is clear that these cancellations will persist for a similar calculation of the one-loop self-energy of the 
LW partner to the Higgs doublet.   When there is no kinetic mixing and the LW higher-derivative term is diagonal ({\em i.e.}, 
when $R_1$ and $O$ are identity matrices) we have a conventional LW gauge theory which is free of quartic and quadratic 
divergences.  Consider how the cancellation of divergences in this limit are modified by Eq.~(\ref{eq:newshift}):  Quartically 
divergent terms from LW loops proportional to $g^T (M^2)^{-1} g$ now proportional to $g^T R_1 O^T (M^2)^{-1} O R_1 g$;  
quadratically divergent terms proportional to $g^T g$ now are each proportional to $g^T  R_1 R_1^T g$  if they originate from 
loops of ordinary particles or $g^T  R_1 O^T O R_1^T g$ if they originate from loops of LW particles.  However, since $O$ is 
orthogonal, the same cancellations occur term by term.  All other divergences cancel as in the ordinary LWSM.

\section{Vertex Displacement}
\begin{figure}[t]
\includegraphics[width = 0.495\textwidth]{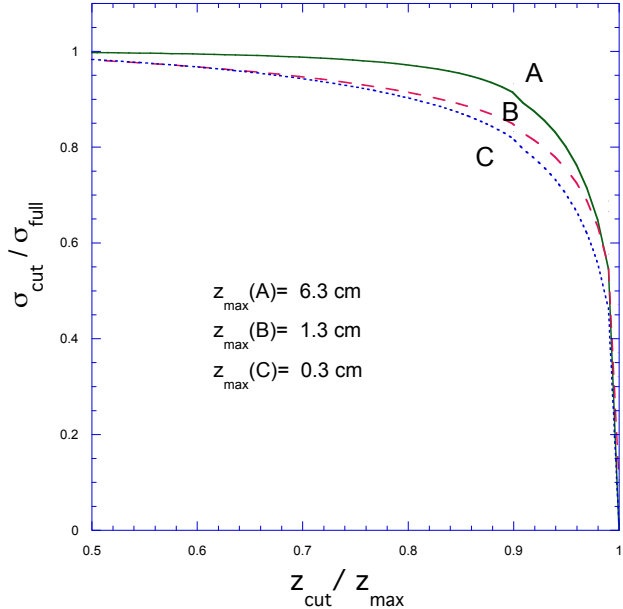}
\caption{\label{fig:two}  LW dark photon production cross section as a function of the wrong displacement vertex cut, for the
three scenarios defined in the text. }
\end{figure}  

One means of experimentally distinguishing an ordinary resonance from a LW resonance has been discussed by
\'{A}lvarez, {\em et al.}~\cite{schat}, based on the study of scattering amplitudes by Grinstein, O'Connell and Wise~\cite{emergent}.   
Unlike an ordinary resonance, the position of the production and decay vertices of a LW resonance are arranged so that 
the LW decay products move toward the primary production vertex rather than away from it.   Hence in an inertial frame in 
which the LW resonance is produced at rest, the time at which the LW resonance is produced is {\em earlier} than the time 
at which the incoming particles collide.  

It must be stressed, however, that this picture follows from an $S$-matrix calculation in which measurements are made only on 
asymptotic wave packet states.  The vertex positions are extrapolated from the four-momenta and positions of the asymptotic 
wave-packets.  Such an extrapolation leads to the apparent acausal relation between vertices described above.    Let us denote
the vertex separation by the four-vector $w^\mu$.  As in Ref.~\cite{schat}, we will assume that the $S$-matrix analysis is valid, {\em i.e.},
that all measurements are made at distances large compared to $|\vec{w}|$ and at times large compared to $w^0$.

One can then use the existence of ``wrong displaced vertices" to place cuts on experimental data in order to isolate 
the signal of LW particles.  For example, in their study of LHC LW signals, \'{A}lvarez, {\em et al.} require the transverse vertex 
displacement to be greater than $20$ microns~\cite{schat}.  Here we note that an analagous approach can be applied in lower-energy
experiments.  As an example, let's consider the production of LW dark photons at low-energy fixed-target experiments.   

Since we have established that a dark photon and its LW partner have the same couplings to ordinary matter, they also have
the same production cross section.  Hence the bounds on dark photons that do not explicitly require a displaced vertex (for 
example, searches for resonant bumps) provide the same bounds on a dark LW photon of the same mass.  Given that the production 
vertex must occur within the fixed target and the LW decay products must move toward it, this implies that the LW decay vertex 
could appear {\em in front of} a suitably thin target rather than behind it.   An important question is whether a cut on the LW resonance 
momentum, to guarantee a measurable vertex separation, substantially depletes the production rate.  We do not find this to be the case.  
To check this, we consider the bremsstrahlung process $e^- Z \rightarrow e^- \tilde{A} \, Z$, where $Z$ represents the atomic number of 
the target nucleus and $\tilde{A}$ is the dark LW photon.  The maximum possible vertex displacement from the target parallel to the
beam direction, $z_{max}$, is determined by the beam energy $E_0$; we can then determine what fraction of the total cross 
section remains if we require $z_{cut} < z < z_{max}$.   We use the differential cross section $d\sigma/(dx \,d\!\cos\theta)$ given 
in the Weiz\"{a}cker-Williams approximation by Eq.~(5) of Ref.~\cite{best},   where the dark LW photon energy $E_{\tilde{A}}= x \,E_0$ and 
the scattering angle $\theta$ is measured with respect to the beam direction.  This formula is valid for $x \,\theta^2 \ll 1$, so in the case 
where $z_{cut}=0$, we require $\theta < \theta_0=5$~degrees. The results don't depend strongly on this choice since the 
differential cross section falls off quickly at angles of order $(m_A/E_0)^{3/2}$ which are typically much smaller.  In the presence 
of the cut, we integrate the differential cross section subject to the constraints
\begin{equation}
x > \frac{1}{E_0} \left[ m_{\tilde{A}}^2 + \frac{z^2_{cut}} {z^2_{max}\cos^2\theta} (E_0^2-m_{\tilde{A}}^2 )  \right]^{1/2} \,\, ,
\label{eq:con1}
\end{equation}
\begin{equation}
\theta < \mbox{min }\left\{\cos^{-1} \left(\frac{z_{cut}}{z_{max}}\right)\,\, , \,\, \theta_0 \right\}\,\, .
\label{eq:con2}
\end{equation}
Eq.~(\ref{eq:con1}) follows from  the requirement that $\beta \gamma \cdot c \tau \cos\theta > z_{cut}$, while Eq.~(\ref{eq:con2}) follows from 
Eq.~(\ref{eq:con1}) and $x<1$.  Results from this integration are shown in Fig.~\ref{fig:two}, for three of the benchmark scenarios described in 
Ref.~\cite{best}:  Scenario A corresponds to $m_{\tilde{A}} = 50$~MeV and $E_0=0.2$~GeV and $\epsilon=1 \times 10^{-5}$; Scenario B 
corresponds to $m_{\tilde{A}} = 200$~MeV and $E_0=6$~GeV and $\epsilon=3 \times 10^{-5}$; Scenario C corresponds to 
$m_{\tilde{A}} = 50$~MeV and $E_0=1$~GeV and $\epsilon=1 \times 10^{-4}$.   The results indicate that the suppression in the cross 
section is not substantial for a wide range of $z_{cut}$.  This is consistent with the qualitative observation in Ref.~\cite{best} that the 
differential cross section is peaked at forward angles and when the dark particle carries nearly the entire beam energy.  As a result, the 
effect of the cut is relatively mild.   

Nevertheless, measurements of such vertex separations at a fixed-target experiment may be challenging.  In the previous examples, 
the $\tilde{A}$ decays in front of the target and its decay products move directly toward the target with a small opening angle, 
of order $m_{\tilde{A}}/E_0$.  If the target obstructs the path between the decay vertex and the point of observation downstream, one
might worry about the effect of further interactions within the target.  This issue is not relevant for thin targets, and can be avoided for thick targets
if they are very narrow (of order the beam size) in at least one transverse dimension.  Moreover, in these examples, vertex resolution 
would have to be typically much better than a centimeter.  These observations suggest that the target and detector engineering would have 
to be optimized to search for the peculiar type of signal expected in these scenarios.

\section{Conclusions}

We have considered the possibility that Lee-Wick partners of dark sector particles might occur at lower
energies than the Lee-Wick partners of standard model particles.  We have studied a theory in which a dark photon
and its Lee-Wick partner have kinetic mixing with hypercharge, including terms of quadratic and quartic order in
derivatives.  We isolated the relevant couplings of the dark Lee-Wick photon to standard model particles.  We verified
that quadratic divergences are cancelled in this theory, just as in garden-variety Lee-Wick theories.  Finally, we
have noted that searches for ``wrong displaced vertices" in low-energy fixed-target experiments are theoretically well motivated.

Although we have focused here on Lee-Wick dark photons, it is possible for other kinds of Lee-Wick particles (for 
example, scalars) to communicate with the visible sector through other portals (for example, the Higgs portal).  These 
alternatives, yet to be studied, may lead to cleaner experimental signals of the physics of interest.  Hence, the present work 
represents an initial exploration of the ways in which a dark sector may provide new avenues for detecting 
the uniquely acausal signatures of Lee-Wick particles.

\begin{acknowledgments}  
The author thanks Todd Averett,  Marc Sher and Josh Erlich for useful comments. This work was supported by the NSF under Grant PHY-1068008.  
\end{acknowledgments}


\end{document}